\documentclass{article}

\usepackage{arxiv}

\usepackage[utf8]{inputenc} % allow utf-8 input
\usepackage[T1]{fontenc}    % use 8-bit T1 fonts
\usepackage{hyperref}       % hyperlinks
\usepackage{url}            % simple URL typesetting
\usepackage{booktabs}       % professional-quality tables
\usepackage{amsmath}
\usepackage{amssymb}
\usepackage{nicefrac}       % compact symbols for 1/2, etc.
\usepackage{microtype}      % microtypography
\usepackage{lipsum}
\usepackage{graphicx}% Include figure files
\usepackage[caption=false]{subfig}
\usepackage{dcolumn}% Align table columns on decimal point
\usepackage{bm}% bold math
\usepackage{microtype}
\usepackage[capitalize]{cleveref}
\usepackage{listings}
\usepackage[mathlines]{lineno}% Enable numbering of text and display math
\newcommand{\bvec}[1]{\mathbf{#1}}
\newcommand{\fb}[1]{\ensuremath{\mathrm{FB}_#1}}
\newcommand{\fbb}[1]{\ensuremath{\bvec{\mathrm{\textbf{FB}}_#1}}}
\newcommand{\kap}{\kappa}
\newcommand{\bet}{\beta}
\newcommand{\Gam}{\bvec{\Gamma}}
\newcommand{\GamT}{\Gam^T}
\newcommand{\gam}{\vec{\gamma}}
\newcommand{\vnu}{\vec{\nu}}
\newcommand{\vx}{\vec{x}}
\newcommand{\vxs}{\vx^*}
\newcommand{\cth}{\cos \theta}
\newcommand{\sth}{\sin \theta}
\newcommand{\cph}{\cos \phi}
\newcommand{\sph}{\sin \phi}

\newcommand{\sthsq}{\sin^2 \theta}
\newcommand{\cphsq}{\cos^2 \phi}
\newcommand{\sphsq}{\sin^2 \phi}
\newcommand{\cfour}{c_4(\kap, \bet)}
\newcommand{\cfive}{c_5(\kap, \bet)}
\newcommand{\csix}{c_6(\kap, \bet, \eta)}
\newcommand{\ceight}{c_8(\kap, \bet, \eta, \vnu)}

\newcommand{\feightexpspherical}{\kap (\nu_1 \cth + \nu_2 \sth \cph + \nu_3 \sth \sph) + \bet \sthsq (\cphsq -\eta \sphsq)}
\newcommand{\intomega}{\int_0^\pi \int_0^{2\pi}}
\newcommand{\dth}{\mathrm{d}\theta}
\newcommand{\dph}{\mathrm{d}\phi}
\newcommand{\domega}{\dph \sth \dth}
\newcommand{\htfo}{{}_2 F_1}
\newcommand{\htfofn}[4]{\htfo \left(#1, #2; #3; #4 \right)}
\newcommand{\hofo}{{}_1 F_1}
\newcommand{\hofofn}[3]{\hofo \left(#1; #2; #3 \right)}
\newcommand{\hzfo}{{}_0 F_1}
\newcommand{\hzfofn}[2]{\hzfo \left(; #1; #2 \right)}
\newcommand{\gammafn}[1]{\Gamma \left(#1 \right)}
\newcommand{\alkj}{a_{l,k,j}}
\newcommand{\bigalkj}{A_{L,K,J}}
\newcommand{\blkj}{B_{L,K,J}}

\title{The 8-parameter Fisher-Bingham distribution on the sphere}% Force line breaks with \\
\author{
  Tianlu Yuan \\
  Dept.~of Physics and Wisconsin IceCube Particle Astrophysics Center\\
  University of Wisconsin\\
  Madison, WI 53706 \\
  \texttt{tyuan@icecube.wisc.edu}
  }

\begin{document}
% \author{Tianlu Yuan}
% \email{tyuan@icecube.wisc.edu}
% \affiliation{Dept.~of Physics and Wisconsin IceCube Particle Astrophysics Center, University of Wisconsin, Madison, WI 53706, USA}

%\date{May 29, 2019}% It is always \today, today,

\maketitle

\begin{abstract}
The Fisher-Bingham distribution (\fb{8}) is an eight-parameter family of probability density functions (PDF) on $S^2$ that, under certain conditions, reduce to spherical analogues of bivariate normal PDFs. Due to difficulties in computing its overall normalization constant, applications have been mainly restricted to subclasses of \fb{8}, such as the Kent (\fb{5}) or von Mises-Fisher (vMF) distributions. However, these subclasses often do not adequately describe directional data that are not symmetric along great circles. The normalizing constant of \fb{8} can be numerically integrated, and recently Kume and Sei showed that it can be computed using an adjusted holonomic gradient method. Both approaches, however, can be computationally expensive. In this paper, I show that the normalization of \fb{8} can be expressed as an infinite sum consisting of hypergeometric functions, similar to that of the \fb{5}. This allows the normalization to be computed under summation with adequate stopping conditions. I then fit the \fb{8} to a synthetic dataset using a maximum-likelihood approach and show its improvements over a fit with the more restrictive \fb{5} distribution.
\end{abstract}


\section{Introduction} \label{sec:intro}

\begin{figure*}[htp!]
\centering
    \subfloat{
        \includegraphics[width=0.49\textwidth]{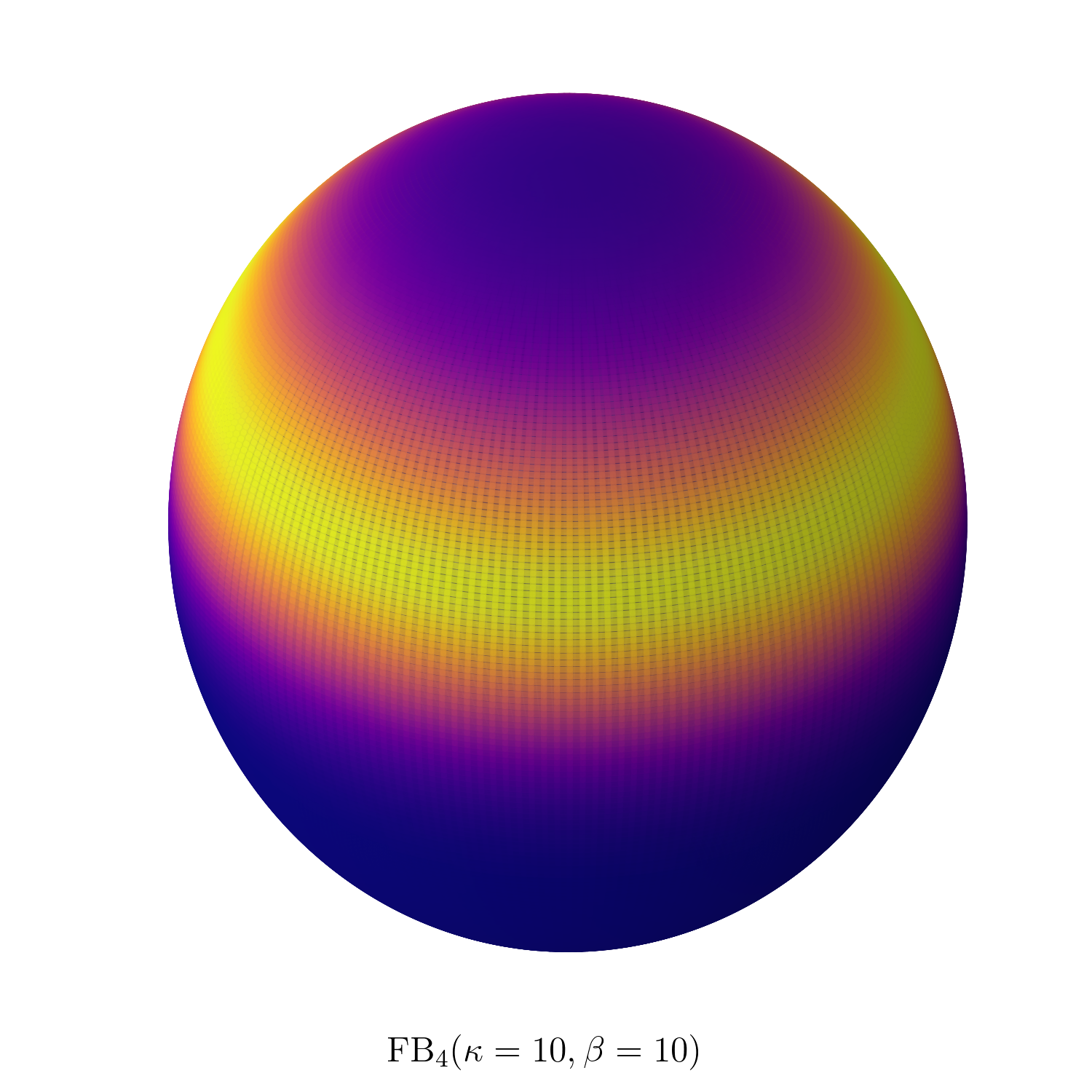}
    }
    \subfloat{
        \includegraphics[width=0.49\textwidth]{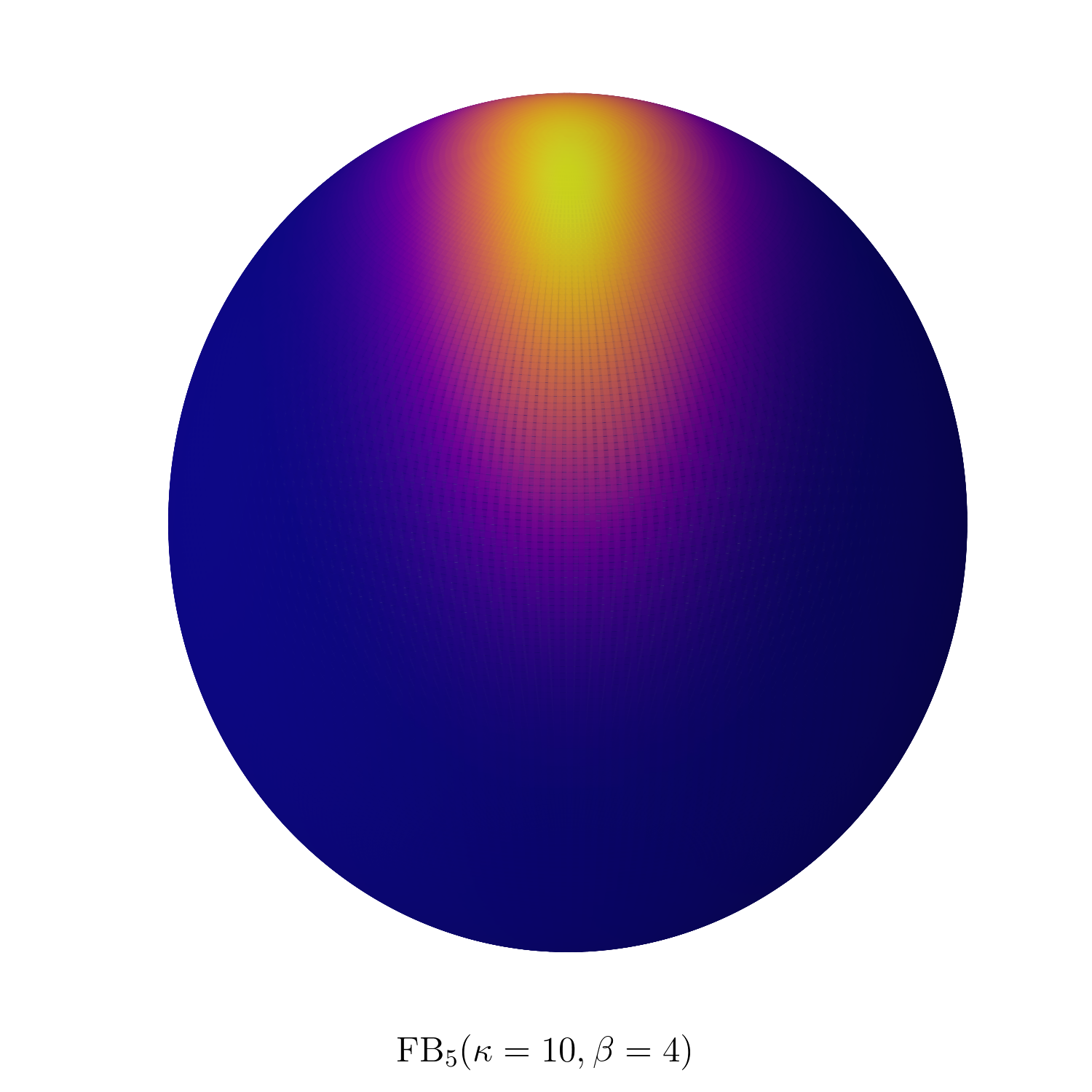}
    }

	\protect\caption{Illustrations of a \fb{4} (left) and a \fb{5} (right) PDF on the sphere. The color map is proportional to the probability density, with brighter regions corresponding to higher densities.}
    \label{fig:fb4fb5}
\end{figure*}

Directional statistics involves the study of probability density functions (PDF) with supports on $S^{n} \subset \mathbb{R}^{n+1}$. This paper will focus on distributions on the sphere, $S^2$. Such distributions have found applications in fields as varied as earthquake modeling to paleomagnetism of lava flows to reconstruction of radio pulses~\cite{mardia2000,fraenkel2014}. A simple and commonly used PDF on $S^2$ that is the analogue to an isotropically distributed, bivariate normal distribution is the von Mises-Fisher (vMF) distribution~\cite{fisher1953}. A more general distribution that is the analogue to a general bivariate normal distribution is the Kent (\fb{5}) distribution~\cite{kent1982},
\begin{equation}\label{eq:fb5}
    f_5(\vx) = \cfive^{-1}\exp \left\{\kap \gam_1 \cdot \vx + \bet[(\gam_2 \cdot \vx)^2-(\gam_3 \cdot \vx)^2]\right\},
\end{equation}
where $\cfive$ is the normalization constant, $\vx$ is a unit vector on $S^2$, $\kap$ and $\bet$ are non-negative parameters, and $\gam_i$ are unit vectors that correspond to the columns of a $3\times 3$ orthogonal matrix, $\Gam$, which determines the orientation of the PDF. An additional constraint, $\kap > 2 \bet$, is required to interpret the \fb{5} distribution as an analogue of the general bivariate normal distribution~\cite{kent1982}, as shown in the right panel of \cref{fig:fb4fb5}. The vMF distribution corresponds to the trivial case of $\bet=0$. Both the vMF and \fb{5} distributions have been well-studied, and have found use in many applications. However, \fb{5} suffers from the restriction that its PDFs must be symmetric across two great circles intersecting at $90^\circ$ on the sphere. Data that clusters along small circles, for example the non-equatorial lines of latitude, are ill-described by \fb{5}.

An alternative to \fb{5} is the small-circle distribution proposed in~\cite{bm4}. This is a four-parameter subclass (\fb{4}) of the Fisher-Bingham family and can be written as,
\begin{align}\label{eq:bm4}
    f_4(\vx) &= \cfour^{-1}\exp \left\{\kap \gam_1 \cdot \vx + \bet[(\gam_2 \cdot \vx)^2+(\gam_3 \cdot \vx)^2]\right\} \\
    &= \cfour^{-1}\exp \left\{\kap \gam_1 \cdot \vx + \bet[1 - (\gam_1 \cdot \vx)^2]\right\}.
\end{align}
With $\kap < 2 \bet$, $f_4$ describes small-circle distributions on the sphere as shown in the left panel of \cref{fig:fb4fb5}. Removing the constraint on $\kap$ and $\bet$, the only difference between \fb{5} and \fb{4} is the sign of the term in square-brackets. The \fb{4} distribution is completely specified by four parameters, since $\gam_1$ is a unit vector. However, it is only a good description of data that is evenly distributed along a small circle. Generalizations are needed in order to model data that falls between the extremes described by \fb{4} and \fb{5}.

In order to perform a maximum likelihood fit of directional data, the PDFs need to be normalized. It was shown in~\cite{bm4} that $\cfour$ can be written in terms of the confluent hypergeometric function. It was shown in~\cite{kent1982} that $\cfive$ can be written as an infinite sum consisting of modified Bessel functions of the first kind. These approaches motivated the calculation of the \fb{8} normalization discussed in \cref{sec:c8}, but first a natural generalization of \fb{4} and \fb{5} is given in \cref{sec:general}.

\section{The Fisher-Bingham distribution} \label{sec:general}

\begin{figure*}[htp!]
\centering
    \subfloat{
        \includegraphics[width=0.49\textwidth]{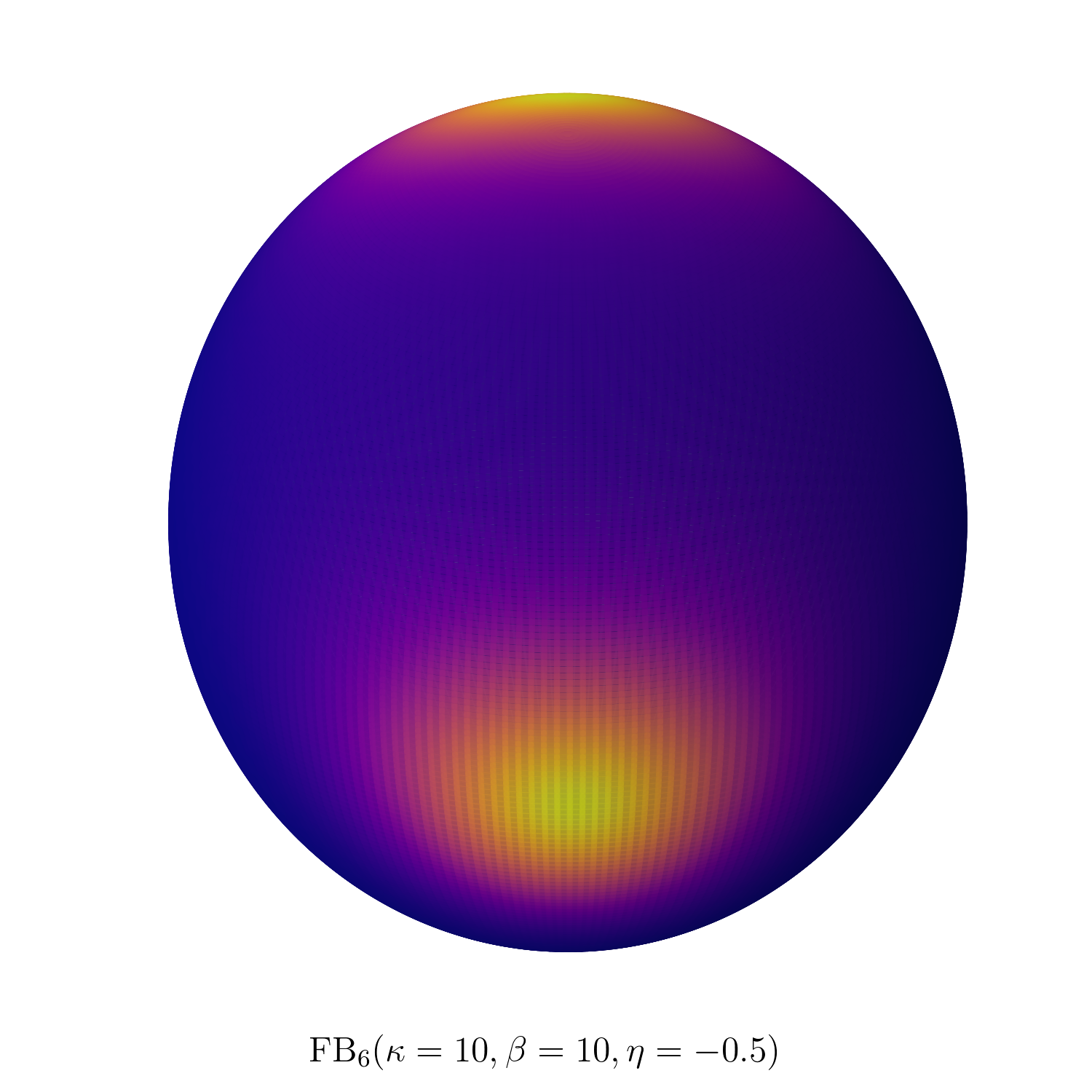}
    }
    \subfloat{
        \includegraphics[width=0.49\textwidth]{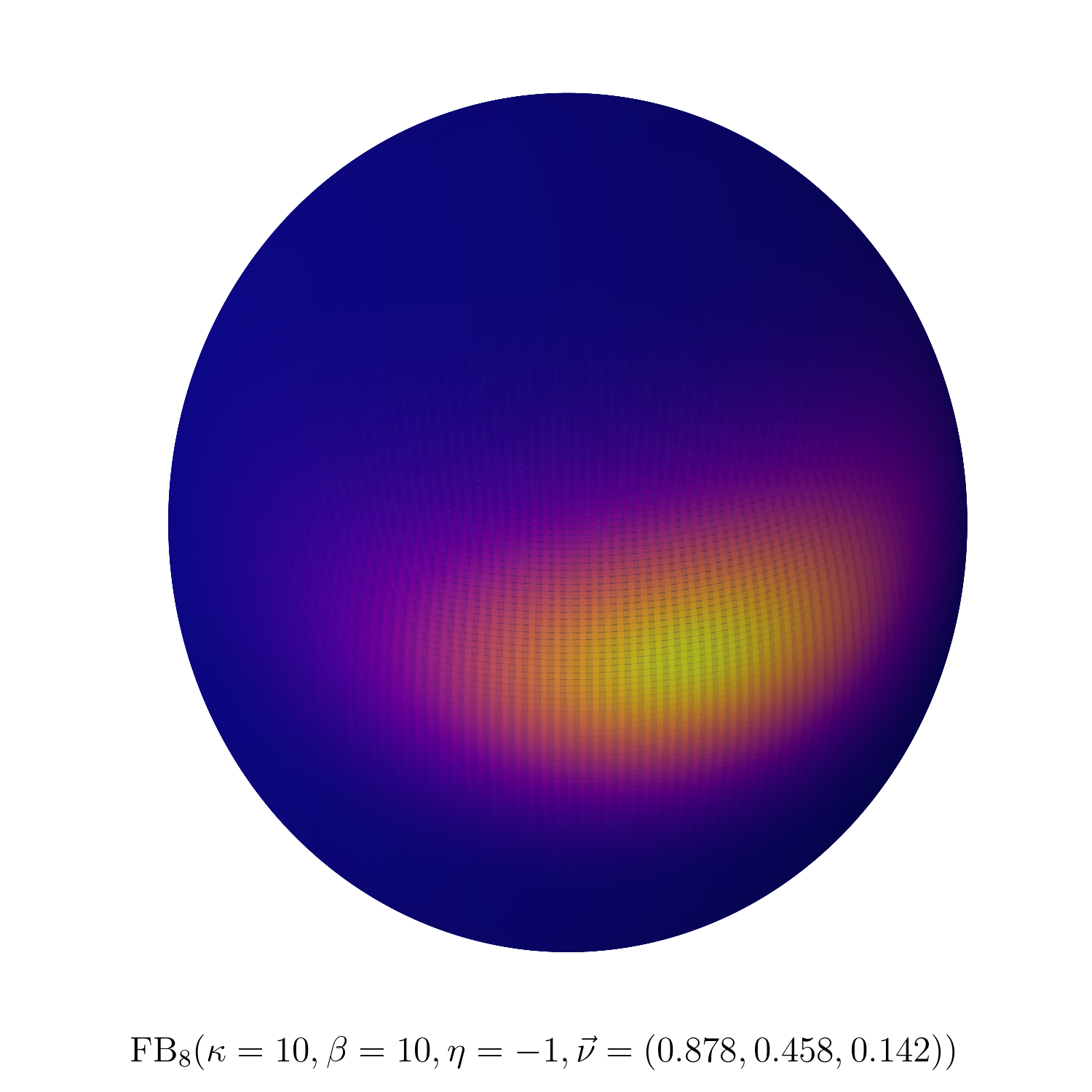}
    }

	\protect\caption{Illustrations of a \fb{6} (left) and a \fb{8} (right) PDF on the sphere. The color map is proportional to the probability density, with brighter regions corresponding to higher densities.}
    \label{fig:fb6fb8}
\end{figure*}

It is simple to construct a 6-parameter PDF (\fb{6}) that generalizes \fb{4} and \fb{5}~\cite{rivest1984}. This is given here as
\begin{align}\label{eq:fb6}
    f_6(\vx) &= \csix^{-1}\exp \left\{\kap \gam_1 \cdot \vx + \bet[(\gam_2 \cdot \vx)^2-\eta(\gam_3 \cdot \vx)^2]\right\},
\end{align}
with $\kap \geq 0$, $\bet \geq 0$ and $|\eta| \leq 1$. Clearly, $\eta=1$ corresponds to \cref{eq:fb5} and $\eta=-1$ to \cref{eq:bm4}. For $\kap > 2 \bet$, $f_6$ has a single maximum, corresponding to where $\vx$ is aligned with $\gam_1$. For $\kap < 2 \bet$, $f_6$ can describe either the small-circle distribution of~\cite{bm4} or a bimodal distribution where the modes are $180^\circ$ degrees apart on a small circle as shown in the left panel of \cref{fig:fb6fb8}. As is the case for \fb{5}, the \fb{6} distribution is symmetric, which means that it cannot describe distributions that lie along small circles with a unique mode.

In order to describe unimodal distributions that lie along small circles, \cite{kim2019} proposed a distribution that is a natural combination of the vMF and \fb{4} distributions. This can be generalized further by combining the vMF and \fb{6} distributions, which results in the Fisher-Bingham distribution (\fb{8}). It is parametrized here as,
\begin{align}
    f_8(\vx)&= \ceight^{-1}\exp \left\{\kap (\Gam \vnu - \gam_1) \cdot \vx \right\}\exp \left\{\kap \gam_1 \cdot \vx + \bet[(\gam_2 \cdot \vx)^2-\eta(\gam_3 \cdot \vx)^2]\right\} \label{eq:fb8exp} \\
    &= \ceight^{-1}\exp \left\{\kap \vnu \cdot \GamT \vx + \bet[(\gam_2 \cdot \vx)^2-\eta(\gam_3 \cdot \vx)^2]\right\}, \label{eq:fb8}
\end{align}
where $\vnu$ is a unit vector on the sphere, and an example is shown in the right panel of \cref{fig:fb6fb8}. The first term in \cref{eq:fb8exp} is proportional to a vMF distribution with mean direction aligned along $\Gam \vnu - \gam_1$, and the second term is proportional to \cref{eq:fb6}. Thus, \fb{8} does not necessarily have to be symmetric about a great circle. When $\vnu = (1,0,0)$ \cref{eq:fb8} reduces to \cref{eq:fb6}.

\section{Calculating the \texorpdfstring{\fbb{6}}{FB6} and \texorpdfstring{\fbb{8}}{FB8} normalizations}\label{sec:c8}

\subsection{Exact series solution}
The \fb{8} distribution was proposed in~\cite{mardia1972, mardia1975}, though it has not been widely applied due to difficulties in computing $\ceight$~\cite{mardia2000}. An exact calculation involving holonomic functions was given in~\cite{Kume2018}, which requires solving ordinary differential equations with the Runge-Kutta method. It can also be estimated using numerical integration. Both of these methods, however, can be computationally expensive. A faster approximation given in~\cite{Kume2005} relies on the saddlepoint method, but this is known to be inexact~\cite{Kume2018}. A fast and accurate calculation of $\ceight$ is desirable to perform maximum likelihood inference using \cref{eq:fb8}. This is the subject of this \namecref{sec:c8} and \cref{sec:num}.

Since $\Gam$ simply enacts a rotation of the sphere, the normalization constants above are independent of $\Gam$ and it is simpler to work in the standard frame, where the coordinate axes are defined by the columns of $\Gam$ with the $z$-axis corresponding to $\gam_1$~\cite{Kume2005}. The coordinate transformation $\vxs = \GamT \vx$ allows us to write \cref{eq:fb8} as,
\begin{equation}\label{eq:fb8standard}
    f_8(\vxs) = \ceight^{-1}\exp \left\{\kap \vnu \cdot \vxs + \bet(x_2^{*2}-\eta x_3^{*2})\right\}.
\end{equation}
In spherical coordinates
\begin{equation}\label{eq:spherical}
    \vxs = (\cth, \sth \cph, \sth \sph)
\end{equation}
and
\begin{equation}\label{eq:fb8spherical}
    f_8(\vxs) = \frac{\exp \left\{ \feightexpspherical \right\}}{\ceight} ,
\end{equation}
where
\begin{align}\label{eq:c8spherical}
    \ceight &= \intomega e^{\feightexpspherical} \domega \\
        % &= \intomega e^{\kap \nu_1 \cth}e^{\kap \nu_2 \sth \cph}e^{\kap \nu_3 \sth \sph}e^{\bet \sthsq [\cphsq - \eta \sphsq]} \domega \\
        &\equiv \intomega \mathcal{I} \dph \dth.
\end{align}
Taylor expanding $\mathcal{I}$ gives,
\begin{align}\label{eq:i8taylor}
     \mathcal{I} &= e^{\kap \nu_1 \cth} \sum_{l,k,j=0}^{\infty} \bigg\{\frac{(\kap \nu_2 \sth \cph)^l}{l!}\frac{(\kap \nu_3 \sth \sph)^k}{k!} \frac{[\bet \sthsq (\cphsq-\eta \sphsq)]^j}{j!}\bigg \} \\ \label{eq:i8taylori}
      %& \quad \times \sum_{i=0}^j (-\eta)^i \binom{j}{i} \cos^{2(j-i)} \phi  \sin^{2i} \phi \sth \bigg\} \\ \label{eq:i8taylori}
%
      %&= e^{\kap \nu_1 \cth} \sum_{l,k,j=0}^{\infty} \bigg\{\frac{(\kap \nu_2 \sth \cph)^l}{l!}\frac{(\kap \nu_3 \sth \sph)^k}{k!} \frac{(\bet \sthsq)^j}{j!} \nonumber \\
      %& \quad \times \sum_{i=0}^j (-\eta)^i \binom{j}{i} \cos^{2(j-i)} \phi  \sin^{2i} \phi \sth \bigg\} \\ \label{eq:i8taylori}
%
     &= e^{\kap \nu_1 \cth} \sum_{l,k,j=0}^{\infty} \sum_{i=0}^j \bigg\{ \frac{\kap^{l+k} \bet^j \nu_2^l \nu_3^k (-\eta)^i}{l! k! j!} \binom{j}{i} \sin^{2j+l+k+1} \theta \cos^{2(j-i)+l} \phi \sin^{2i+k} \phi \bigg \}.
\end{align}
The integration proceeds as in~\cite{kent1982} by applying Eq.~(6.2.1) and Eq.~(9.6.18) in~\cite{abramowitz_stegun}. Noting that the integral over $\phi$ vanishes unless $k$ and $l$ are both even,
\begin{align}
     \ceight &= 2 \int_0^\pi \sum_{l,k,j=0}^{\infty} \bigg\{ \frac{\kap^{2(l+k)} \bet^j \nu_2^{2l} \nu_3^{2k}}{(2l)! (2k)! j!}  e^{\kap \nu_1 \cth} \sin^{2(j+l+k)+1} \theta \nonumber \\
     &\quad \times \sum_{i=0}^j (-\eta)^i \binom{j}{i} B\left(j-i+l+\frac{1}{2}, i+k+\frac{1}{2}\right)\bigg\} \dth \\
     &= 2 \sqrt{\pi} \sum_{l,k,j=0}^{\infty} \bigg\{ \frac{\kap^{2(l+k)} \bet^j \nu_2^{2l} \nu_3^{2k}}{(2l)! (2k)! j!}\left | \frac{\kap \nu_1}{2}\right |^{-j-l-k-\frac{1}{2}} I_{j+l+k+\frac{1}{2}}(|\kap \nu_1|) \nonumber \\
     &\quad \times \sum_{i=0}^j (-\eta)^i \binom{j}{i} \gammafn{j-i+l+\frac{1}{2}} \gammafn{i+k+\frac{1}{2}} \bigg\} \\
%     
    %  &= 2 \sqrt{\pi} \sum_{l,k,j=0}^{\infty} \bigg\{ \frac{\kap^{2(l+k)} \bet^j \nu_2^{2l} \nu_3^{2k}}{(2l)! (2k)! j!}\left | \frac{\kap \nu_1}{2}\right |^{-j-l-k-\frac{1}{2}} I_{j+l+k+\frac{1}{2}}(|\kap \nu_1|) \nonumber \\
    %  &\hspace{4em} \gammfn{k+\frac{1}{2}} \gammafn{j+l+\frac{1}{2} } \htfofn \left(-j, k+\frac{1}{2}; \frac{1}{2}-j-l; -\eta \right) \bigg\} \nonumber \\
%     
     &= 2 \sqrt{\pi} \sum_{l,k,j=0}^{\infty} \bigg\{ \frac{\kap^{2(l+k)} \bet^j \nu_2^{2l} \nu_3^{2k}}{(2l)! (2k)! j!} \frac{\gammafn{k+1/2} \gammafn{j+l+1/2}}{\gammafn{j+l+k+3/2}} \nonumber \\
     &\quad \times \hzfofn{j+l+k+\frac{3}{2}}{\frac{\kap^2 \nu_1^2}{4}} \htfofn{-j}{k+\frac{1}{2}}{\frac{1}{2}-j-l}{-\eta} \bigg\} \label{eq:c8series}
\end{align}
where $\Gamma$ denotes the gamma function, $B$ the beta function, $I_v$ the modified Bessel function of the first kind, $\hzfo$ the confluent hypergeometric limit function, and $\htfo$ the Gaussian hypergeometric function. The last equality can be derived from Eq.~(9.6.47) and Eq.~(15.4.1) in~\cite{abramowitz_stegun}. These special functions can be evaluated numerically, and $\ceight$ can be computed to good approximation with adequate stopping conditions on $j$, $k$, and $l$. By setting $\vnu = (1,0,0)$, the only nonzero terms occur when $k=l=0$ and \cref{eq:c8series} simplifies to
\begin{equation}\label{eq:c6series}
    \csix = 2 \pi \sum_{j=0}^{\infty} \bigg\{ \frac{\bet^j \gammafn{j+1/2}}{j! \gammafn{j+3/2}} \hzfofn{j+\frac{3}{2}}{\frac{\kap^2}{4}} \htfofn{-j}{\frac{1}{2}}{\frac{1}{2}-j}{-\eta} \bigg\},
\end{equation}
the normalization for \fb{6}. Further setting $\eta=1$ and applying Eq.~(15.1.21) in~\cite{abramowitz_stegun} recovers $\cfive$ as computed in~\cite{kent1982}.

\subsection{Closed-form approximation for \texorpdfstring{$\bvec{\csix}$}{c6}}

If $\kap$ or $\bet$ is large, $\csix$ can be approximated piecewise in two separate regimes: $\kap < 2 \bet$ and $\kap > 2 \bet$. For $\kap < 2 \bet$, note that we can rewrite
\begin{align}
    \csix &= \intomega e^{\kap \cth + \bet \sthsq (\cphsq -\eta \sphsq)} \domega \\
    &= e^{\bet \left(1+\frac{\kap^2}{4\bet^2}\right)}\intomega e^{-\bet \left(\cth - \frac{\kap}{2\bet}\right)^2} e^{-\bet(1+\eta)\sthsq \sphsq} \domega \\
    &= e^{\bet \left(1+\frac{\kap^2}{4\bet^2}\right)}\int_{-1}^1 \int_0^{2\pi}  e^{-\bet \left(z - \frac{\kap}{2\bet}\right)^2} e^{-\bet(1+\eta)(1-z^2) \sphsq} \mathrm{d}\phi \mathrm{d}z,
\end{align}
where $z=\cth$. In order to factorize the integration over $\phi$ and $z$, we make the assumption that the term modulated by $\sphsq$ can be fixed to $z = \kap/(2 \bet)$, where the maximum of $f_6$ occurs in latitude. Then, using Eq.~(13.1.27) in~\cite{abramowitz_stegun},
\begin{align}
    \csix &\approx e^{\bet \left(1+\frac{\kap^2}{4\bet^2}\right)}\int_{-1}^1 \int_0^{2\pi}  e^{-\bet \left(z - \frac{\kap}{2\bet}\right)^2} e^{-\bet(1+\eta)\left(1-\frac{\kap^2}{4 \bet^2} \right) \sphsq} \mathrm{d}\phi \mathrm{d}z \\
        &\approx 2 \pi  e^{\bet \left(1+\frac{\kap^2}{4\bet^2}\right)} \hofofn{\frac{1}{2}}{1}{\bet (1+\eta) \left(\frac{\kap^2}{4 \bet^2} -1\right)}  \int_{-1}^1  e^{-\bet \left(z - \frac{\kap}{2\bet}\right)^2}  \mathrm{d}z \\
        &\approx 2 \pi  e^{\bet \left(1+\frac{\kap^2}{4\bet^2}\right)} \hofofn{\frac{1}{2}}{1}{\bet (1+\eta) \left(\frac{\kap^2}{4 \bet^2}-1\right)} \sqrt{\frac{\pi}{\bet}} \qquad (\kap < 2\bet) \label{eq:c6approxb}.
\end{align}
The last line uses an approximation for large $\beta$~\cite{bm4}.

In the case of $\kap > 2 \bet$, $f_6$ is maximal at $\theta=0$ and for large $\kap$ becomes approximately a bivariate normal distribution. By setting $\phi=0$ and $\phi = \pi/2$ and Taylor expanding in $\theta$, we see that
\begin{equation}\label{eq:c6approxk}
    \csix \approx 2 \pi e^\kap [(\kap - 2 \bet) (\kap + 2 \bet \eta)]^{-\frac{1}{2}} \qquad (\kap > 2\bet), 
\end{equation}
which is similar to Eq.~(3.5) in~\cite{kent1982}. A comparison of \cref{eq:c6series} to \cref{eq:c6approxb,eq:c6approxk} is shown in the right panel of \cref{fig:lnc} (dashed blue). This approximation is accurate away from $\kap=2\bet$. The saddlepoint approximation of~\cite{Kume2005} is also shown (dotted orange).

These approximations are useful when working with large $\kap$ or $\bet$, where it may not be possible to numerically compute \cref{eq:c6series} due to extremely large terms in the summand. With maximum likelihood estimation, for example, it is often simpler to work with $\ln \csix$, which can be approximated with \cref{eq:c6approxb,eq:c6approxk} without running into computational overflows. For $\ceight$, unfortunately, no closed-form approximation is known, and the options are to perform numerical integration, use the method proposed in~\cite{Kume2018}, or use \cref{eq:c8series}. The saddlepoint method was tested to be accurate for $\csix$ but not for $\ceight$.

% $\kap=0$. Nevertheless, it is possible to calculate $\ckzero$ directly from the integral form. Again, the only nonzero terms occur when $k=l=0$ such that,
% \begin{equation} \label{eq:ckzero}
%     \ckzero = 4 \pi \sum_{j=0}^{\infty} \frac{\beta^j {}_2F_1(\frac{1}{2}, -j, \frac{1}{2}-j, -\eta)}{(2 j+1) j!}.
% \end{equation}
% Along with the exact solutions for the uniform ($\kap = \bet =0$) and vMF ($\bet = 0$ and $\vnu = (1,0,0)$) distributions, this edge case should be treated separately when computing the normalization.

\section{Numerical computation} \label{sec:num}

The infinite series in \cref{eq:c8series,eq:c6series} can be evaluated by truncation under an appropriate stopping condition. To simplify the notation, let $\alkj$ be the summand of \cref{eq:c8series} such that $\ceight = \sum_{l,k,j=0}^\infty \alkj$. For $\eta \leq 0$, $\alkj$ is non-negative for all $l,k,j$ while for $\eta >0$, $\alkj$ is guaranteed to be non-negative only for even $j$. As such, $\alkj$ may be an alternating sequence in $j$. Furthermore, $|\alkj|$ does not in general decrease monotonically in any of the indices, but does so only after a certain point.

\begin{figure*}[htp!]
\centering
    \subfloat{
        \includegraphics[width=0.49\textwidth]{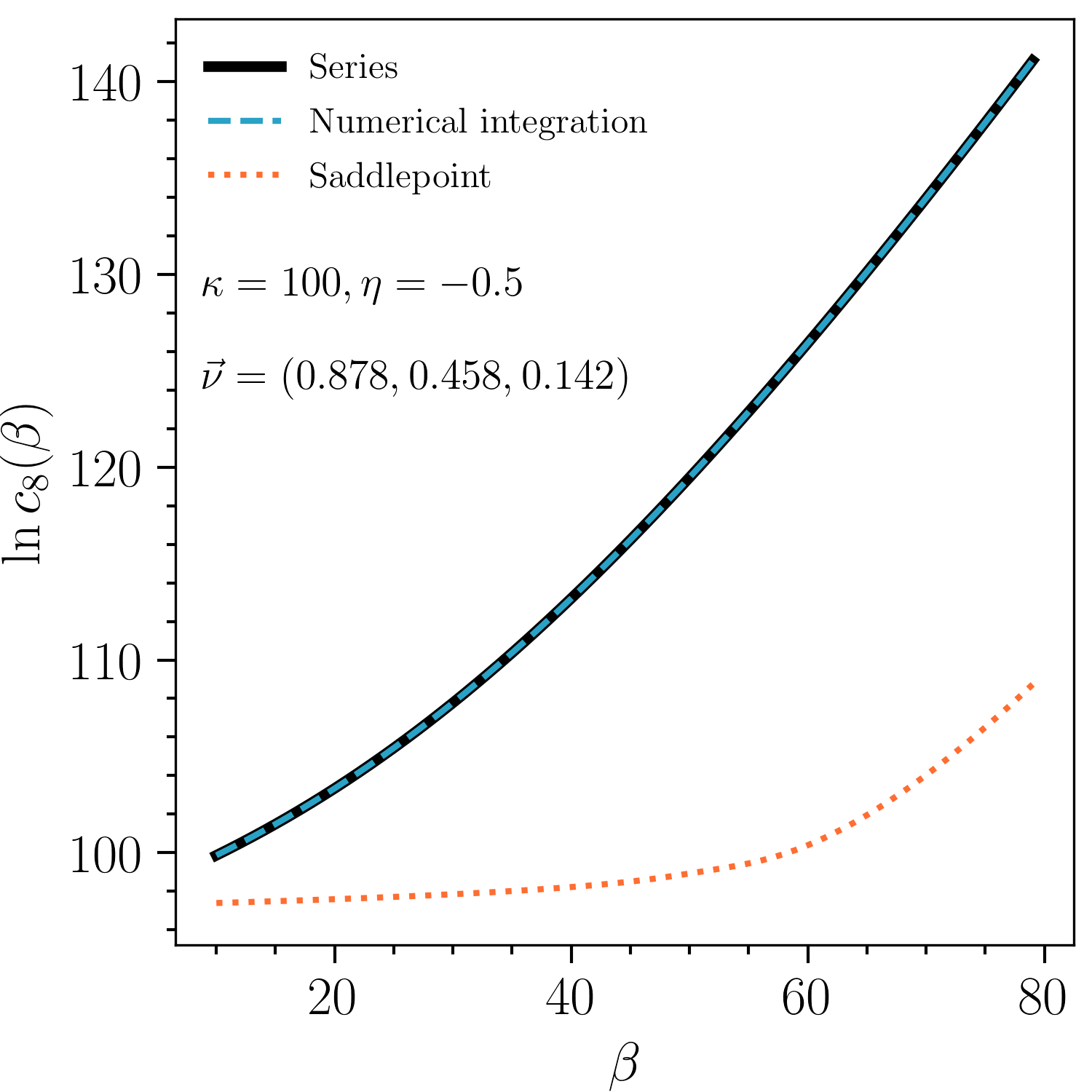}
    }
    \subfloat{
        \includegraphics[width=0.49\textwidth]{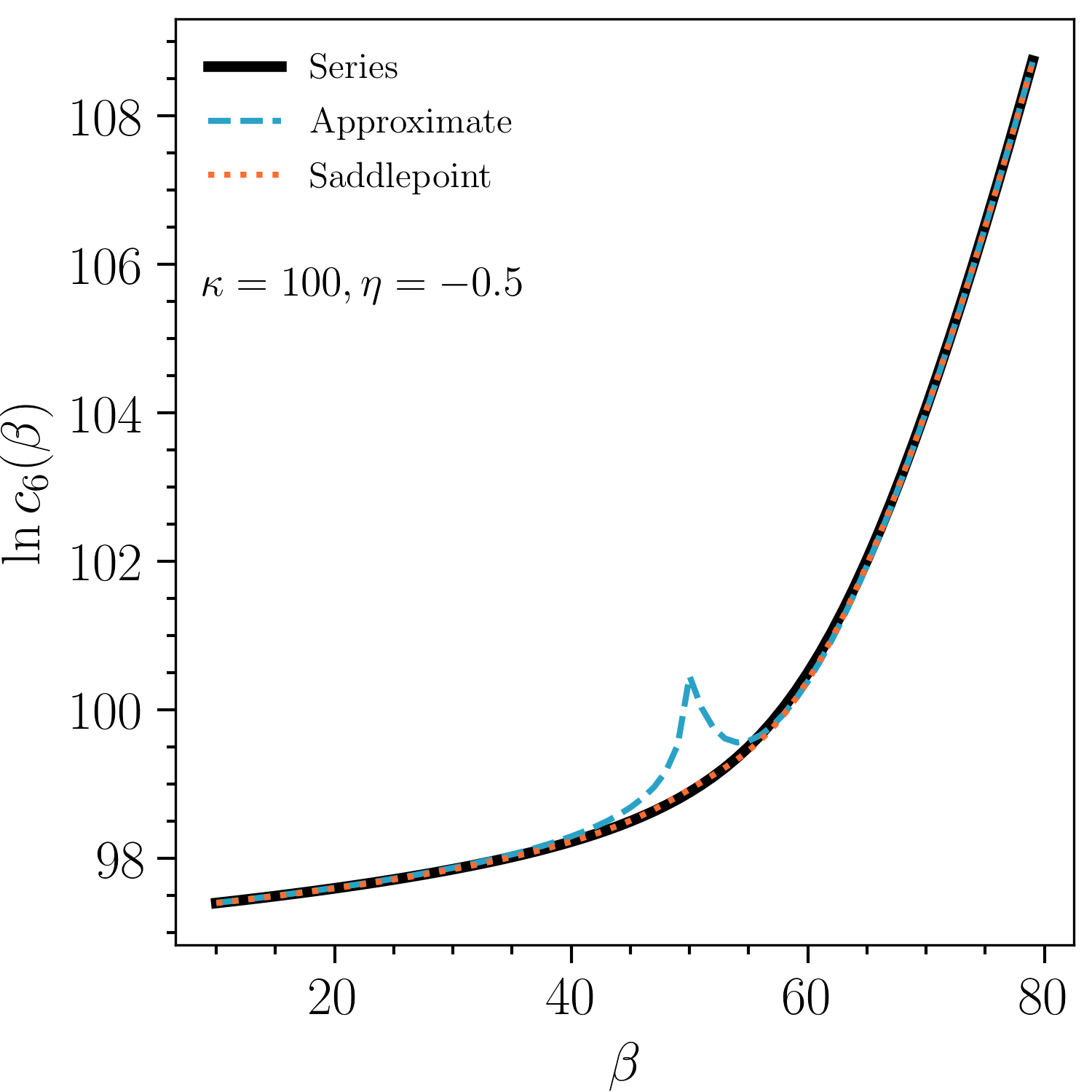}
    }

	\protect\caption{The left panel shows a comparison of $\ln c_8(\beta)$, with remaining parameters fixed, as calculated via the series summation (solid black) and via a numerical integration routine (dashed blue). The right panel shows a comparison of $\ln c_6(\beta)$ as calculated via the series summation (solid black) and using the approximations in \cref{eq:c6approxb,eq:c6approxk} (dashed blue). The saddlepoint approximation of~\cite{Kume2005} is included in both panels (dotted orange) and performs well for $\ln c_6(\beta)$.}
    \label{fig:lnc}
\end{figure*}

The most efficient algorithm to estimate $\ceight$ would be to reindex and order $b_n = \alkj$ such that $|b_{n+1}| \geq |b_n|$ for all $n\in \mathbb{W}$, and perform the summation starting from $n={0,1,2 \ldots}$ up to certain tolerance. However, this ordering is difficult to evaluate. In practice, a robust calculation can be obtained by setting a step size $s \in 2 \mathbb{N}$ and defining 
\begin{align}\label{eq:chunks}
    \bigalkj &\equiv \sum_{l=sL}^{s(L+1)-1} \left[  \sum_{k=sK}^{s(K+1)-1} \left( \sum_{j=sJ}^{s(J+1)-1} \alkj \right) \right]\\
    \blkj &\equiv \sum_{l=sL}^{s(L+1)-1} \left[  \sum_{k=sK}^{s(K+1)-1} \left( \sum_{j=sJ}^{s(J+1)-1} |\alkj | \right) \right].
\end{align}
This ensures that an even number of $\alkj$ terms are summed at each step and $\ceight = \sum_{L,K,J=0}^\infty \bigalkj$. Then for some tolerance $\epsilon >0$, the stopping algorithm is
\begin{lstlisting}[mathescape=true,language=Python]
$L$, $P_L$, $\tilde{c}_8$ = 0
while True:
    $K$, $P_{L,K}$, $S_L$ = 0
    while True:
        $J$, $P_{L,K,J}$, $S_{L,K}$ = 0
        while True:
            $\tilde{c}_8$ += $\bigalkj$
            $S_{L}$ += $\blkj$
            $S_{L, K}$ += $\blkj$
            if $\blkj < |\tilde{c}_8|\epsilon$ and $\blkj \leq P_{L,K,J}$:
                break
            $P_{L,K,J}$ = $\blkj$
            $J$ += 1
        if $S_{L, K} < |\tilde{c}_8|\epsilon$ and $S_{L, K} \leq P_{L,K}$:
            break
        $P_{L,K}$ = $S_{L, K}$
        $K$ += 1
    if $S_L < |\tilde{c}_8|\epsilon$ and $S_L \leq P_L$:
        break
    $P_L$ = $S_L$
    $L$ += 1
\end{lstlisting}
where $\tilde{c}_8$ denotes the series calculation of $\ceight$. This nested summation routine loops through $J$, $K$, and $L$ in that order and breaks once the contribution to the partial sum of $\blkj$ for the current index is within the tolerance and is less than the previous term. In tests, $\epsilon=10^{-12}$ and $s=13$ seemed to perform well. A comparison of $\ln c_8$ computed using the series summation and a numerical integration routine (QUADPACK) is shown in the left panel of \cref{fig:lnc}.

The evaluation of \cref{eq:c6series} follows the same procedure as above, but with a single summation over $J$ while setting $l=k=0$. This amounts to just running the innermost loop. One final thing to note is that $\alkj$ often contains large terms in its numerator and denominator that may often cancel each other. Computationally it may be difficult to evaluate them separately, and a better approach is to work with $\exp(\ln \alkj)$, explicitly taking the logarithm before exponentiation.

\section{Example application}\label{sec:example}

To illustrate the performance of the \fb{8} distribution in modeling directional data, a synthetic dataset was randomly sampled from an \fb{8} distribution that peaked along a small circle on the sphere. An unbinned maximum likelihood fit was then performed using the \fb{5} and \fb{8} distributions~\cite{kent1982, mardia2000}. The SLSQP routine was used to perform a constrained fit of \fb{5} and the L-BFGS-B routine was used for a bounded fit of \fb{8}~\cite{NoceWrig06}.

\begin{figure*}[htp!]
\centering
    \subfloat{
        \includegraphics[width=0.49\textwidth]{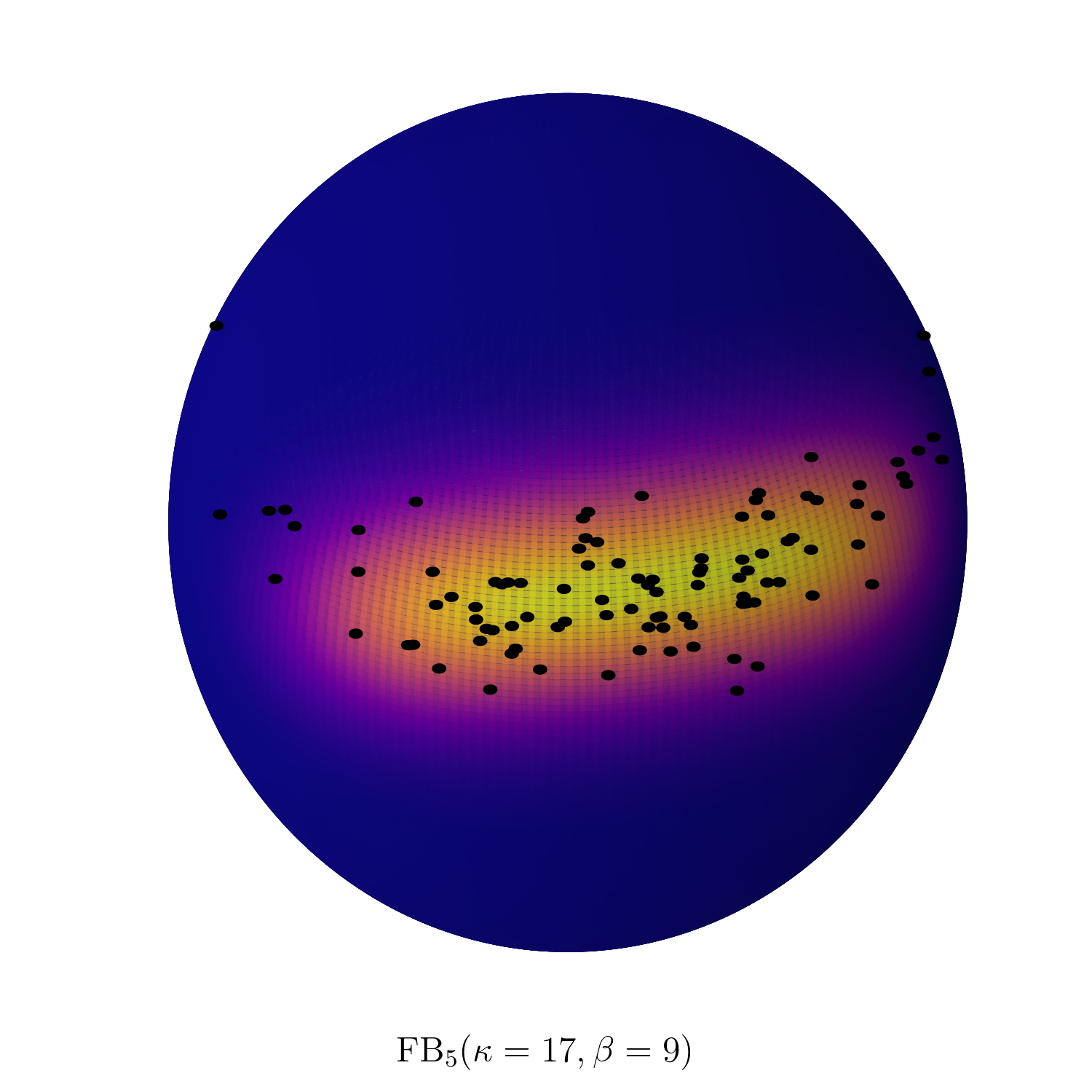}
    }
    \subfloat{
        \includegraphics[width=0.49\textwidth]{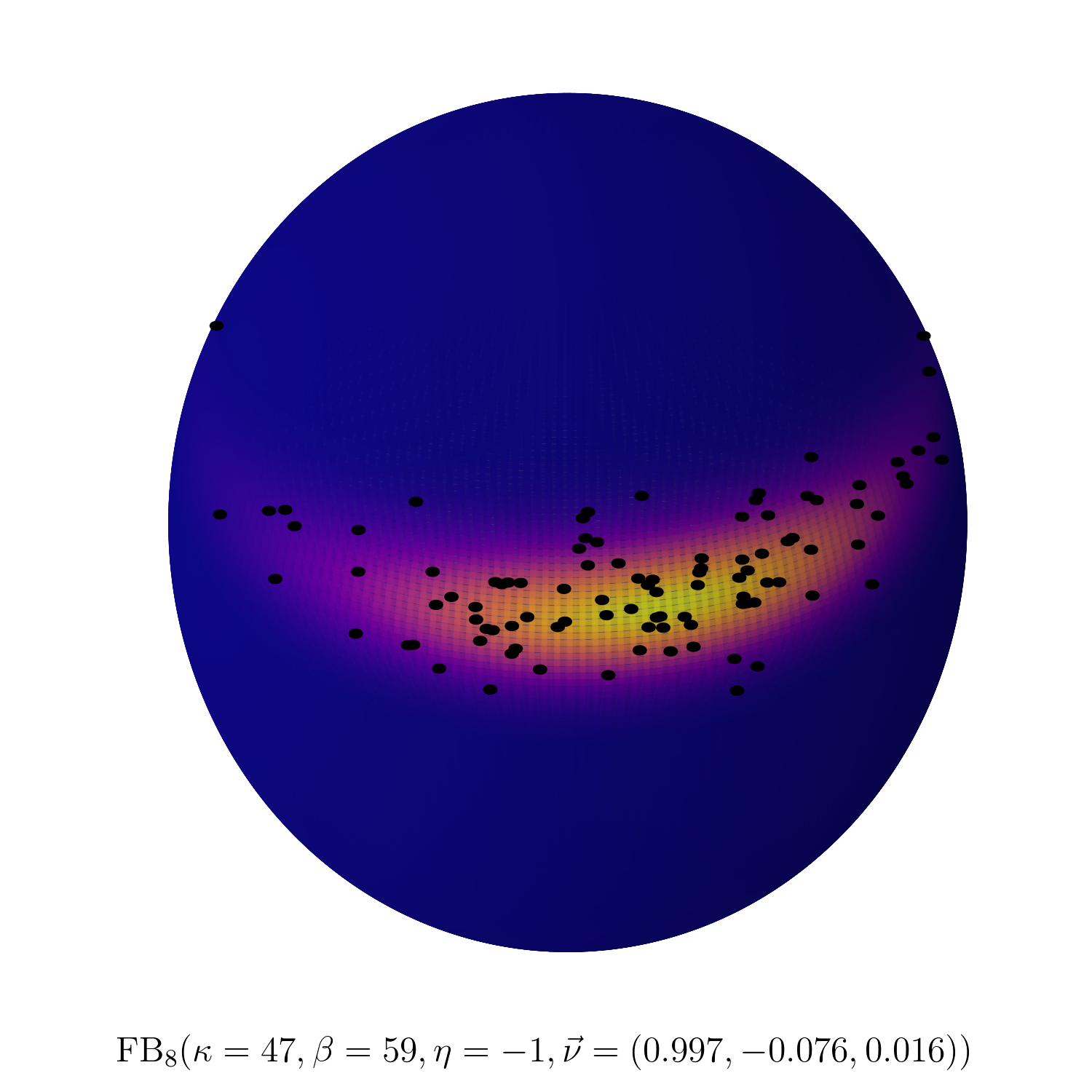}
    }
	\protect\caption{Maximum likelihood fit of synthetic dataset (black points) using \fb{5} (left) and \fb{8} (right). The best-fit parameters for each distribution are indicated in the underlying text. The data is better described by the \fb{8} distribution.}
    \label{fig:toy}
\end{figure*}

The results are shown in \cref{fig:toy}. The synthetic data (black points) is poorly described by the \fb{5} distribution, but well described by \fb{8}. The best-fit negative log-likelihoods are $31.7$ for the \fb{5} and $-12.4$ for the \fb{8}, indicating much better data agreement using \fb{8}.

\section{Conclusion} \label{sec:conclusion}

In this paper, I have calculated the normalization of the 8-parameter Fisher-Bingham distribution on $S^2$ using its series expansion. This is given in \cref{eq:c8series}. By construction, the normalization for the \fb{6} distribution~\cite{rivest1984} is a simplification and given in \cref{eq:c6series}. Further, a piecewise approximation of $\csix$ was derived in closed form and seems to perform well for large $\kap$ or $\bet$, away from the region where $\kap = 2 \bet$.

An algorithm for computing \cref{eq:c8series} numerically was described in \cref{sec:num}. As the sequence of $\alkj$ is not, in general, non-negative and only decreases in absolute value to zero after a certain point, a truncation tolerance based on successive partial sums is not sufficient to robustly calculate the normalization. Instead, the proposed technique groups contiguous $\alkj$ into $\bigalkj$ and $\blkj$ as defined in \cref{eq:chunks}. The stopping condition is then described using partial sums of $\blkj$. The series calculation of the normalization is shown to be robust and matches that obtained from numerical integration. The summation can be computationally much faster than numerical integration, although this depends on their respective tolerance settings.

With $\ceight$ in hand, exact maximum likelihood fits can be performed using \cref{eq:fb8}. As an example, a synthetic dataset was generated along a small circle on the sphere. Maximum likelihood fits performed using \fb{5} ill-described the data, while fits using \fb{8} exhibited better agreement. As \fb{8} is a superset of \fb{5}, it should allow for more flexible descriptions of directional data. A \texttt{Python} package, extended from the implementation of \fb{5} in~\cite{fraenkel2014}, is available and contains all the distributions described in this paper\footnote{https://github.com/tianluyuan/sphere}.

\section*{Acknowledgements}

This work stemmed out of discussions with Dmitry Chirkin and attempts to fit directional data from~\cite{Chirkin2013}. I would also like to thank Kareem Farrag and Austin Schneider for discussions on other potential approaches to calculate the normalization. The author is supported in part by NSF grant PHY-1607644 and by the University of Wisconsin Research Committee with funds granted by the Wisconsin Alumni Research Foundation.

\bibliographystyle{unsrt}
\bibliography{main}

\end{document}